\documentclass{aa}
\usepackage[varg]{txfonts}
\usepackage{natbib}
\usepackage{graphicx}
\bibpunct{(}{)}{;}{a}{}{,} % to follow the A&A style
\usepackage{color}

\begin{document}

\title{The VISTA\thanks{Based on observations made with ESO Telescopes at the La Silla Paranal Observatory under program ID 088.C-0117} ~Carina Nebula Survey}
\subtitle{II. Spatial distribution of the infrared-excess-selected young stellar population}

\author{P.~Zeidler\inst{\ref{inst1},\ref{inst2}} \and T.~Preibisch\inst{\ref{inst1}} \and T.~Ratzka\inst{\ref{inst1},\ref{inst3}}\and V.~Roccatagliata\inst{\ref{inst1}} \and M.~G.~Petr-Gotzens\inst{\ref{inst4}}}

\institute{Universit\"ats-Sternwarte M\"unchen, Ludwig-Maximilians-Universit\"at, Scheinerstr. 1, 81679 M\"unchen, Germany\label{inst1} \and Astronomisches Rechen-Institut, Zentrum f\"ur Astronomie der Universit\"at Heidelberg, M\"onchhofstr. 12-14, 69120 Heidelberg, Germany\label{inst2} \and Institute for Physics/IGAM, NAWI Graz, Karl-Franzens-Universit\"at, Universit\"atsplatz 5/II, 8010 Graz, Austria\label{inst3} \and European Southern Observatory, Karl-Schwarzschild-Str. 2, 85748 Garching, Germany\label{inst4}}

\date{Received date /
Accepted date }

\abstract{We performed a deep wide-field (6.76~square-degrees) near-infrared survey with the VISTA telescope that covers the entire extent of the Carina nebula complex (CNC). The point-source catalog created from these data contains  around four~million individual objects down to masses of 0.1~$M_\odot$.
We present a statistical study of the large-scale spatial distribution and an investigation of the clustering properties of infrared-excesses objects, which are used to trace disk-bearing young stellar objects (YSOs). A selection based on a near-infrared ($J-H$) versus ($H-K_s$) color-color diagram shows an almost uniform distribution over the entire observed area. We interpret this as a result of the very high degree of background contamination that arises from the Carina Nebula's location close to the Galactic plane. Complementing the VISTA near-infrared catalog with \textit{Spitzer} IRAC mid-infrared photometry improves the situation of the background contamination considerably. We find that a ($J-H$) versus ($K_s-[4.5]$) color-color diagram is well suited to tracing the population of YSO-candidates (cYSOs) by their infrared excess. We identify 8\,781 sources with strong infrared excess, which we consider as cYSOs. This sample is used to investigate the spatial distribution of the cYSOs with a nearest-neighbor analysis. The surface density distribution of cYSOs agrees well with the shape of the clouds as seen in our \textit{Herschel} far-infrared survey. The strong decline in the surface density of excess sources outside the area of the clouds supports the hypothesis that our excess-selected sample consists predominantly of cYSOs with a low level of background contamination. This analysis allows us to identify 14 groups of cYSOs outside the central area.Our results suggest that the total population of cYSOs in the CNC comprises about 164\,000 objects, with a substantial fraction ($\sim35\%$) located in the northern, still not well studied parts. Our cluster analysis suggests that roughly half of the cYSOs constitute a non-clustered, dispersed population.}

\keywords{Stars: formation -- Stars: low-mass --ISM: individual objects: NGC3372 -- Stars: pre-main sequence -- Infrared: stars} 

\authorrunning{Zeidler~et~al.} 
\titlerunning{VISTA Carina Survey II: The infrared-excess-selected population}
\maketitle

\section{Introduction}
The Carina nebula complex \citep[CNC, hereafter; see][for an overview]{Smith2008Carina} is one of the largest, most massive, and most active star-forming complexes in our Galaxy. Located at a very well known distance of 2.3~kpc \citep{Smith2006OB}, the nebulosities and the corresponding clouds, as observed in \textit{Herschel} far-infrared (FIR) observations \citep{Preibisch2012Herschel1}, show an overall linear extension of about 2.5~degrees (corresponding to physical scales of about 100~pc) and cover a total area of almost three square degrees on the sky. Since this area is much larger than the typical field of view (FOV) of most optical and infrared cameras at modern telescopes most observational studies of the Carina nebula obtained in the past focused on the central region of the nebula, which harbors the prominent young stellar clusters Tr~14, 15, and 16. Although studies of the central area provided important insight into the stellar populations and the consequences of the massive stellar feedback on the remaining clouds \citep[see, e.g.,][]{Smith2010South_Pillars}, they did not allow the entire CNC and its properties to be assessed.

The spatial extension of  several recent surveys of the CNC provided new information about the global properties of the region. These include the deep near-infrared (NIR) survey of the central $\sim~0.36$~square-degree region with HAWK-I at the ESO 8m-VLT \citep{Preibisch2011bNIR_XRAY,Preibisch2011cDisks}, the 1.4~square-degree X-ray survey performed in the \textit{Chandra} Carina Complex Project \citep[CCCP; see ][for an overview]{Townsley2011CCCP}, and a \textit{Spitzer} MIR study of the CCCP field \citep{Povich2011b_extinction}. The cloud structure in the central $\sim1.6$~square-degree area has been studied in our LABOCA/APEX sub-mm survey \citep{Preibisch2011cDisks}, and we have recently used the \textit{Herschel} FIR observatory to perform a survey of about a five square-degree area that covers the entire extent of the CNC \citep{Preibisch2012Herschel1,Roccatagliata2013Herschel3}. With the available \textit{Spitzer} and \textit{Herschel} observations, the overall structure of the clouds in the CNC has now been investigated well over its full spatial extent. The \textit{Herschel} observations also revealed several hundred protostellar objects \citep{Gaczkowski2012Herschel2}, which are dispersed over the full spatial extent of the cloud complex.\\
On the other hand, sensitive studies of the YSO populations were still restricted to the central regions, basically the CCCP field, and even there only the $\sim0.36$~square-degree region covered by HAWK-I (i.e., just $\sim0.36/2.5$ or 14~\% of the full area of the cloud complex) had been observed with sufficient sensitivity to allow the detection of YSOs down to $0.1~M_\odot$. This lack of deep NIR data that cover the full spatial extent of the CNC was the motivation for the new very wide-field survey we have performed with the ESO 4m Visible and Infrared Survey Telescope for Astronomy \citep[VISTA; ][]{Emerson2006VISTA}.

In the VISTA Carina Nebula Survey (VCNS), we observed a 2$\times$2 tile mosaic covering an area of 6.67~deg$^2$ ($2.30\,{\rm deg}\,\times\,2.94\,{\rm deg}$) in the NIR $J$-,$H$-, and $K_s$-bands with the VIRCAM camera \citep[see][]{Dalton2006VIRCAM}. The details of the observations, the data reduction, and the photometric calibration are described in \citet{Preibisch2013Carina_a}. This survey revealed 4\,840\,807 individual NIR
sources, of which 3\,951\,580 were reliably detected in at least two bands. The $5\sigma$ magnitude limits are $J_{5\sigma} \approx 20.0$~mag, $H_{5\sigma} \approx 19.4$~mag, and $K_{s,5\sigma} \approx 18.5$~mag, while the effective completeness limits are $J_{\text{compl}} \approx 18.2$~mag, $H_{\text{compl}} \approx 17.5$~mag, and $K_{s,\text{compl}} \approx 17.1$~mag. The photometry was calibrated to the 2MASS system, and the calibration uncertainties were found to be $\sigma$($J, H, K_s) \approx [0.05-0.06]$~mag. Considering the pre-main-sequence (PMS) evolutionary models of \citet{Baraffe1998evolution_model} and a typical age of $\sim3$~Myr, stars with masses as low as $\sim0.1~M_\odot$ are clearly detected in all three bands. At the $5\sigma$ magnitude limits, even brown dwarfs with masses as low as $0.03~M_\odot$ can be detected.

The VCNS source catalog is very well suited to studying the spatial distribution of the young stellar population, since at ages of a few Myr, a considerable number of the YSOs is expected to have circumstellar material, which can be detected by its characteristic infrared excess emission. The exponential decay time of near- to mid-infrared excesses has been established to be $\sim5$~Myr \citep{Fedele_2010}, and thus YSOs with ages up to $\sim5$ -- 10~Myr can be detected by infrared-excess emission. The young stellar clusters in the central parts of the Carina nebula have ages of $\lesssim1$~Myr for the “Treasure Chest” cluster, $\sim1$ -- 2~Myr for the cluster Tr~14, $\sim3$~Myr for Tr~16, and $\sim5$ -- 8~Myr for Tr~15 \citep[see discussion in ][]{Preibisch2011cDisks}. At these ages, considerable portions of the young stellar populations should still show detectable infrared excesses, and therefore this excess will provide us with important information on the overall population of the young stars in the complex.

We note that the strategy of our study is complementary to those of the previous efforts by \citet{Smith2010South_Pillars} and \citet{Povich_2011a}, who identified individual YSOs in selected (but relatively small) regions of the CNC by a detailed analysis and modeling of their spectral energy distributions. In our study, we do not aim at a reliable identification of individual YSOs. Our goal is to statistically investigate the number and the spatial distribution of the YSOs in the CNC. The young sources are photometrically selected.

In this paper we use the VCNS NIR source catalog in combination with \textit{Spitzer} mid-infrared (MIR) data to identify objects with infrared excesses. In Sect. \ref{sec:NIR_selection} we describe the results of the selection by NIR excesses and show that the contamination by Galactic background objects is too high to obtain useful information on the CNC population. In Sect. \ref{sec:MIR} we describe how we extended the VCNS NIR catalog to the MIR by including photometry from \textit{Spitzer} data. Sect. \ref{sec:spatial_dist} discusses the nearest-neighbor analysis for the determination of the surface density of the young stellar population to derive the spatial extent of the CNC and to identify several new clusters. In Sect. \ref{sec:YSO} we estimate the total size of the young stellar content in the CNC, which is based on a comparison of our infrared results to extrapolations based on the X-ray selected population of young stars. Finally, Sect. \ref{sec:summary} provides a concluding discussion.

\section{Near-infrared excess selection}
\label{sec:NIR_selection}
\begin{figure}
\resizebox{\hsize}{!}{\includegraphics{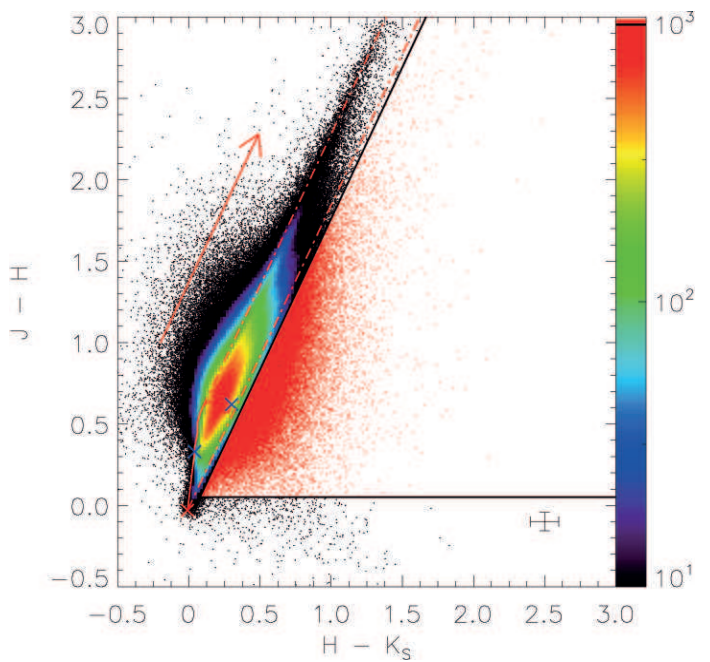}}
\caption{Hess diagram of the $(J-H)$ vs. $(H-K_s)$ color-color diagram. The color indicates the number of objects in each bin with a size of 0.02~mag~$\times$~0.02~mag. A logarithmic scale has been used. The red arrow represents the reddening vector for $A_V=10$~mag based on a slope of 1.86. The solid red line shows the expected colors of main sequence stars according to the model of \citet{Baraffe1998evolution_model}. The blue crosses correspond to $1.0~M_\odot$ and $0.1~M_\odot$ stars, the red cross indicates the position of A0 stars. The dashed-dotted red lines show the reddening band for main sequence stars. The black lines represent the infrared excess selection with $J-H>0.05$~mag, $H-K_s>0.05$~mag, and 0.05~mag below and to the right of the reddening band of the main sequence.We also provide the typical uncertainties of the photometry derived in \citet{Preibisch2013Carina_a}.} In total 86\,175 objects are classified as NIR excess sources shown by red dots.
\label{fig:CC_J-H_H-K_05_05}
\end{figure}

In the first step of our analysis, we identify YSO candidates by their excesses in the NIR range. For this we constructed the $(J-H)$ vs. $(H-K_s)$ color-color diagram (CCD). We only consider objects with a S/N $>10$ in all three NIR filter bands, in order to minimize misclassifications due to photometric uncertainties. Figure \ref{fig:CC_J-H_H-K_05_05} shows the CCD of the 2\,388\,895 sources that satisfy the selection criterion.

The infrared excess classification requires knowledge of the slope of the reddening band in the CCD, which is related to the extinction law. The extinction law depends on the grain composition, shape, and size distribution of each individual cloud. The average extinction law for a specific region on the sky is thus the sum of the contributions from all absorbing clouds along this specific line of sight.

\subsection{The slope of the reddening band}

We used the very large number of objects in our NIR CCD (see Fig. \ref{fig:CC_J-H_H-K_05_05}) to draw the observed distribution of reddened main sequence stars. This provides a good empirical constraint for the reddening vector.

The slope of the reddening band can be expressed as a ratio of the NIR color excess $E$ defined by the wavelength-dependent extinction $A$ \citep{Stead2009NIR_exctinction_law}:

\begin{equation}
\frac{E_{J-H}}{E_{H-K_s}}=\frac{A_J-A_H}{A_H-A_{K_s}}.
\end{equation}

The extinction ratios can be described by a power law \citep{Mathis1990Exctinction,Draine2003Dust_Grains}:

\begin{equation}
\frac{A_x}{A_J}=\left(\frac{\lambda_J}{\lambda_x}\right)^\alpha
\end{equation}

\noindent with the respective central wavelengths. Combining these equations gives the relationship between the three NIR bands:

\begin{equation}
\label{eq:slope}
\frac{E_{J-H}}{E_{H-K_s}}=\frac{\left(\frac{\lambda_H}{\lambda_J}\right)^\alpha -1}{1- \left(\frac{\lambda_H}{\lambda_{K_s}}\right)^\alpha}.
\end{equation}

\noindent This equation describes the slope of the reddening vector as a function of $\alpha$, which is variable and depends on the line of sight determined by observations.

We tested several different $\alpha$ parameters obtained by \citet{Rieke_Lebofsky1985EXT_LAW} and \citet{Stead2009NIR_exctinction_law} and looked for which one empirically fits our CCD best (see Fig. \ref{fig:CC_J-H_H-K_05_05}). We found that a value of $\alpha= 2.1$, corresponding to a slope of 1.86, describes the reddening in our CCD well.
Using this slope, we classified all those objects that are found with $J-H > 0.05$~mag, $H-K_s > 0.05$~mag, and more than 0.05~mag below and to the right to the reddening band for main sequence colors (which corresponds to the typical uncertainty of the photometric calibration of the VISTA data; see \citet{Preibisch2013Carina_a}) as NIR excess sources in order to minimize contamination from photometric uncertainty effects.
 
This selection gives a total number of 86\,175 NIR-excess sources, which makes them 2.7\% of all three-band detections and 3.6\% of the sources with S/N>10 in all three bands.

\subsection{Spatial distribution of the NIR excess sources}
\label{subsec:spatial_dist_NIR}

Analyzing the spatial distribution of the NIR excess sources in the VCNS field, we found that although the density of NIR excess sources appears to be slightly higher in the central region of the survey field (i.e., in the inner parts of the Carina nebula), there is an almost uniform distribution across the full extent of the VCNS field. Since we know from the \textit{Herschel} observations that the clouds associated to the CNC cover only the inner part of the 6.7~square degrees of the VCNS field, many of the NIR excess sources are found far ($\gtrsim 10$~pc) outside the boundaries of the clouds. This suggests that the NIR excess sources at the edges of the VCNS field (and thus outside the clouds) are most likely not YSOs, but rather background sources.

From studies of infrared-excess-selected objects in other star forming regions, it has been found that the contamination of the young stellar population is caused by background objects, such as planetary nebulae or evolved Be stars or galaxies \citep{Oliveira_2009,Rebull2010,Preibisch2011cDisks,Koenig2012}. Depending on the celestial position of a specific region, contamination rates of $\sim30\%$ to more than 50\% have been found \citep{Oliveira_2009}. In the case of the Carina nebula, we have to expect a particularly high level of background contamination, because this region is located just at the Galactic plane and, furthermore, close to the the tangent point of the Carina-Sagittarius spiral arm.

In conclusion, the very high number of NIR excess sources (86\,175) and their almost uniform spatial distribution in the VCNS field, in particular outside the area of the clouds of the CNC, show that this NIR excess-selected sample is very strongly contaminated by background sources. It can thus not used for a reliable characterization of the CNC’s overall YSO population.

\section{Complementing the VCNS catalog with the \textit{Spitzer} photometry}
\label{sec:MIR}

Extending the wavelength range to the MIR provides a better basis for identifying young stars by means of their infrared excesses since for YSOs, the MIR contributes strongly to the overall spectrum \citep{Dauphas2011SED,Dullemond2010Protoplanetary_Disks,Lada1984SED,Bodenheimer2011,Prialnik2009_stellar_evolution}. We therefore used data from the \textit{Spitzer} observatory to extend the wavelength range of the VCNS catalog to the MIR range covered by IRAC (3.6 -- $8.0~\mu$m) \citep{Fazio_2004}.

\subsection{The \textit{Spitzer} MIR catalog}

\begin{figure}
\resizebox{\hsize}{!}{\includegraphics{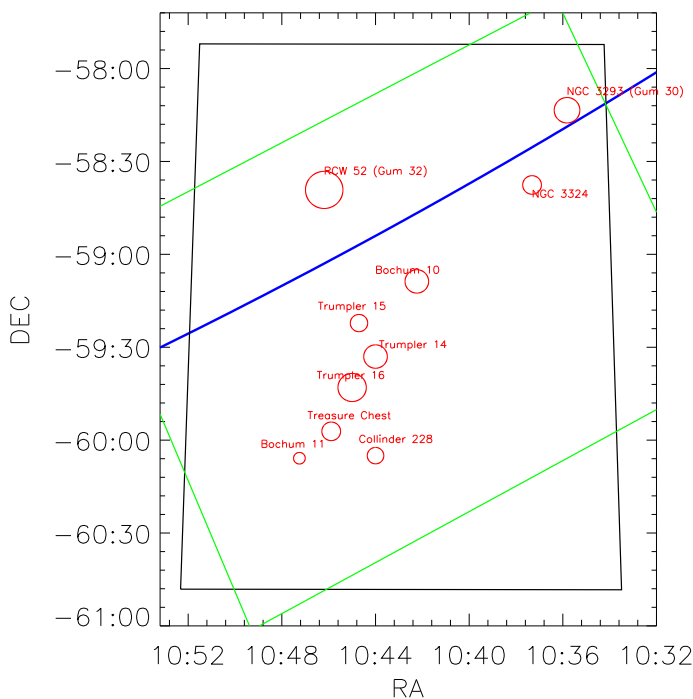}}
\caption{Outline of the VCNS (black) and \textit{Spitzer} (green) survey areas in J2000 coordinates. The blue line indicates the Galactic plane. For orientation purposes, the red circles indicate the locations of the most prominent and well-known clusters in the Carina nebula.}
\label{fig:survey_area}
\end{figure}

The \textit{Spitzer} observatory has performed several observations of the CNC in the context of different surveys during its cold mission phase. When combined, these data cover almost the entire extent of the CNC (see Fig. \ref{fig:survey_area}). Some parts of these \textit{Spitzer} data have
already been analyzed \citep{Smith2010South_Pillars,Povich_2011a}, but no comprehensive study of the full available data set has been published so far.

As described in detail in \citet{Ohlendorf2013Carina}, we have retrieved all available \textit{Spitzer} IRAC data of the CNC from the \textit{Spitzer} Heritage Archive and created a mosaic covering an area of 6.39~deg$^2$. Using the MOsaiker and point source EXtracktor package \citep[MOPEX; ][]{Mopex}, we produced point-source catalogs and extracted the fluxes in all four IRAC bands. These single-band catalogs were then merged to a combined catalog, which is based on the source positions in the $4.5~\mu$m band. This catalog provides fluxes for 237\,947 objects in the IRAC 1 ($3.6~\mu$m) band, 338\,782 objects in the IRAC 2 ($4.5~\mu$m) band, 92\,660 objects in the IRAC 3 ($5.8~\mu$m) band, and 52\,844 objects in the IRAC 4 ($8.0~\mu$m) band. The typical detection limits of these data are 16.1~mag, 15.7~mag, 13.7~mag, and 13.5~mag for the IRAC 1, 2, 3, and 4 bands, respectively. The estimated completeness limits are $\sim13.5$~mag, $\sim14.0$~mag, $\sim12.5$~mag, and $\sim11.8$~mag. Comparing these magnitudes to the pre-main sequence models of \citet{Baraffe1998evolution_model}, we can detect the photospheric emission of stars with masses down to $\sim0.5~M_\odot$ in the IRAC 1 and 2 bands assuming an age of $\sim 1$~Myr. For a more detailed description of this catalog, we refer to \citet{Ohlendorf2013Carina}.

\subsection{The combination of the VCNS NIR catalog with the \textit{Spitzer} MIR catalog}
\label{subsec:combining_VISTA_SPITZER}

Figure \ref{fig:survey_area} shows the areas of the VCNS and \textit{Spitzer} survey. The area covered by both surveys has a size of 5.30~deg$^2$. It comprises 78\% of the VCNS area, and 83\% of the \textit{Spitzer} survey area. The entire extent of the CNC is included in both surveys.

For reliable matching of the \textit{Spitzer} catalog with the VCNS catalog an angular distance of less than $0.5''$ between a \textit{Spitzer} and a VISTA source is required. This angular distance is derived by a well-defined gap between the first and second nearest neighbors in a histogram of the radial distances. This results in 242\,807 matches (6.1\% of all VISTA sources). The numbers of objects with available photometry in the different bands is given in Table \ref{tab:VISTA_sources}. In principle, an analysis based on the VISTA NIR and all four IRAC MIR bands would provide the broadest wavelength coverage and thus allow the best characterization of infrared excesses. However, as can be seen from the source numbers listed in Table \ref{tab:VISTA_sources}, the numbers of detected sources in the IRAC 3 and IRAC 4 bands are much smaller than those detected in the IRAC 2 band. We therefore decided to use the IRAC 2 band for our analysis. This band provides not only the largest number of point source detections, but also has the additional advantage of not being affected by emission from polycyclic aromatic hydrocarbons (PAHs), which can contaminate the three other IRAC bands \citep{Whitney2008IRAC_PAH,Povich_2013}. PAHs have strong lines at $3.3~\mu$m, $6.2~\mu$m, $7.7~\mu$m, and $8.6~\mu$m \citep{Dewangen2009IRAC_PAH}. PAH emission from diffuse structures close to point sources can thus adulterate the photometric flux measurements in the IRAC bands 1, 3, and 4. This could contaminate an excess-selected sample. By using only the photometry of the IRAC 2 (4.5~$\mu$m) band, we aim at minimizing this problematic effect.

\begin{table*}
\caption{Statistics of the numbers of sources in our final photometric catalog, including all sources with S/N<10.}
\label{tab:VISTA_sources}
\centering
\begin{tabular}{crrrr}
\hline
\hline
filter & \multicolumn{1}{c}{central wavelength} & \multicolumn{1}{c}{number of detections} &  \multicolumn{1}{c}{completeness limit} & \multicolumn{1}{c}{detection limit} \\
 & \multicolumn{1}{c}{[$\mu$m]} &  &  \multicolumn{1}{c}{[mag]}  & \multicolumn{1}{c}{[mag]}  \\
\hline
$J$  &  1.25 & 3\,635\,667  &   18.2  & $20.0$ \\
$H$  &  1.65 &  3\,891\,906 & 17.5  & $19.4$ \\
$K_s$  & 2.15  & 3\,568\,837  & 17.1  & $18.5$ \\
 3-band ($J+H+K_s$) &   & 3\,193\,250  &    &  \\ 
 IRAC 1 &  3.6 & 192\,000  &   $13.5$ &  16.1 \\
 IRAC 2 &  4.5 & 242\,807  &    $14.0$ &  15.7 \\
 IRAC 3 &  5.8 & 73\,524  &   $12.5$  &  13.7 \\ 
 IRAC 4 &  8.0 & 37\,544  &   $11.8$  &  13.5 \\ 
 4-band ($J+H+K_s+$IRAC 2) &   & 239\,172  &     &  \\ 
\hline 
\end{tabular}
\end{table*}

\subsection{MIR excess selection}
\label{subsec:MIR_selection}

\begin{figure}
\resizebox{\hsize}{!}{\includegraphics{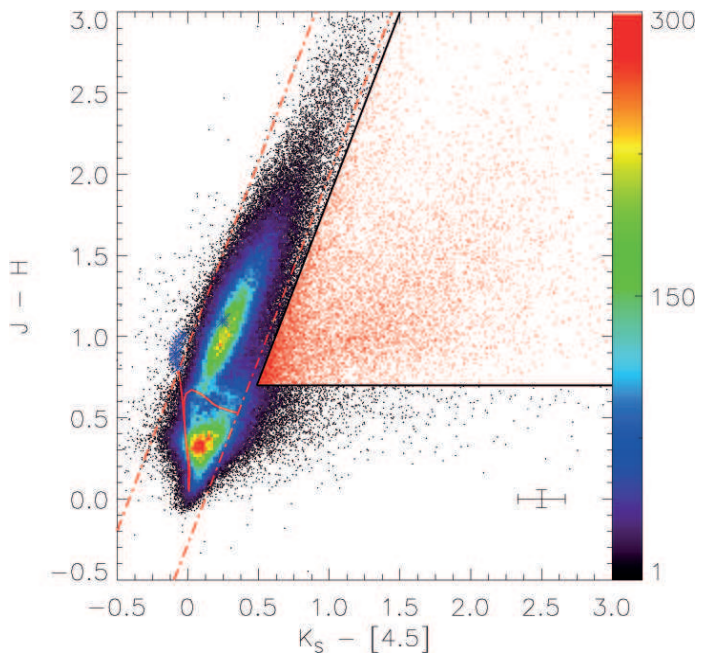}}
\caption{$(J-H)$ vs. $(K_s-[4.5])$ CCD of the 181\,854 sources with SNR$>10$ in all four filter bands. The dashed-dotted red lines show the reddening band for main sequence colors, based on a slope of 2.27 for the extinction
vector. The solid red line shows the main sequence stars from the Padova stellar models \citep{Bressan_2012} and the blue asterisks show the locus of the red supergiants, TP-AGBs, and upper-RGBs produced with the dusty models. We also give the typical uncertainties of the photometry from \citet{Preibisch2013Carina_a} for the VISTA data and from \citet{Ohlendorf2013GUM31} for the [4.5]-band. The color bar shows the number of objects per bin (bin size$=0.02$~mag $\times$ 0.02~mag). The black lines mark the color limits for the infrared excess selection: $(K_s-[4.5] > 0.49)$ and $(J-H) < [(K_s-[4.5])-0.49]\cdot2.27+0.7$. In total 8\,781 objects are classified as infrared excess sources shown by red dots.}
\label{fig:CC_J-H_K-IRAC2_07_45}
\end{figure}

Figure \ref{fig:CC_J-H_K-IRAC2_07_45} shows the $J-H$ vs. $K_s-[4.5]$ CCD diagram of those 181\,854 sources that have photometry with S/N~$>10$ in all four filter bands. The slope of the reddening band can be estimated from the data to be 2.27.

We selected only those sources below the reddening band that satisfy the additional conditions $(K_s-[4.5]) > 0.49$~mag and $(J-H) > 0.7$~mag. This very conservative selection leaves only objects with quite strong infrared excesses and can be expected to provide a cleaner (but necessarily also smaller) sample. The 8\,781 MIR excess objects comprise about 5\% of all sources in the diagram. We denote them as candidate YSOs (cYSOs) in the following text.\\ 

We note that the requirement of a detection in the [4.5]-band removes most of the very faint NIR excess sources ($J\gtrsim18.0$, $H\gtrsim16.5$, $K_s\gtrsim16.0$). Since many (or most) of these are likely background objects, the degree of contamination will automatically be reduced by this step. But this will also remove most of the low-mass ($<0.5~M_\odot$) population in the CNC, leading to an increase in the low-mass limit of the NIR catalog from $0.1~M_\odot$ to $0.5~M_\odot$ for the combined VISTA and \textit{Spitzer} photometric catalog. Our conservative selection criterion also prevents from a mis-selection due to the photometric uncertainties.

The complete infrared excess catalog is only available in electronic form at the CDS\footnote{\url{http://cdsweb.u-strasbg.fr/cgi-bin/qcat?J/A+A/}}. For guidance, a portion of the catalog is given in the Appendix \ref{tab:IR_exces_cat}.

\subsection{Estimation of the background contamination}
The Carina nebula is located close to the Galactic plane, so the background contamination can be considered being quite severe \citep[e.g.,][]{Povich_2011a}. In addition to the high number of field stars, possible contamination can originate in asymptotic giant branch (AGB) stars, background galaxies, supergiants, and the upper red giant branch (RGB) \citep{Robitaille_2008,Oliveira_2009}. Those that are generally very luminous objects can appear with an infrared color excess in our selection imitating YSOs. Another possible sources of contamination are carbon-rich AGB stars (c-AGBs) and extreme AGB stars (x-AGBs). They tend to fall in the infrared excess region but due to their wide range of infrared excess it is difficult to assess the possible numbers \citep{Suh_11}.

Selecting only objects right of the reddening band and applying an additional color cut at $J-H>0.7$~mag, we reduced the number of contaminating background objects that are reddened by the clouds. The possibility that large uncertainties in the photometry may mimic an infrared excess is minimized by our requirement that only objects with a photometric S/N$>10$, resulting in a photometric error $\lesssim 0.1$~mag, (see Subsect. \ref{subsec:MIR_selection}) were considered.

Another potential problem might arise in regions with strong cloud extinction, where particularly strong reddening causes large uncertainties in the colors of stars and might mimic an infrared excess. The column density map constructed from \textit{Herschel} data \citep[see Fig. 6 in][]{Preibisch2012Herschel1} shows that in more than 90\% of the studied area the cloud extinction is $A_V \lesssim 2.5$~mag, and just 1\% exceeds a value of $A_V \sim 5$~mag. Stronger cloud extinction is restricted to a few, mostly rather small cloud clumps, most of which are located near the center of the complex. Therefore, the possible contamination of the photometry in areas of particularly strong cloud extinction does not affect our results on the large-scale spatial distribution of young stars in the complex.

Nevertheless, comparing the locus of where to expect these objects in our CCD, we compared the location of the main sequence with the position for dusty luminous stars \citep{Margio_2008} meaning AGBs, RGBs, and supergiants. Therefore, we used the Padova PARSEC 2.1S model \citep{Bressan_2012}. Most of these objects are primarily located within or left of the main sequence. However, there will be some contamination left. As noted already in Sect. \ref{subsec:spatial_dist_NIR}, this is also shown by other studies \citep{Oliveira_2009,Rebull2010,Preibisch2011cDisks,Koenig2012}, which found that the contamination is caused by background objects, such as planetary nebulae, evolved Be stars or galaxies, and post-AGB stars.

To quantify this possible contamination, \citet{Oliveira_2009} and \citet{Oliveira_10} studied the brightest of the 235 YSOs in the Serpens molecular cloud. They were carefully selected by \citet{Harvey_07}. \citet{Oliveira_2009} used optical spectroscopy of 78 of these sources to find that 25\% of their targets are actually contaminants. \citet{Oliveira_10} studied 147 sources with the Spitzer InfraRed Spectrograph (IRS) and found a contamination
rate of 22\%. They concluded that this is due to the close location to the Galactic plane. In addition, we estimated our contamination with the criteria of \citet{Winston_07}. They also used only \textit{Spitzer} data.
From our infrared-excess-selected sample, we used those objects that are detected in addition in all four IRAC filters. From our 8\,781, a reduced number of 748 objects remain, and 682 objects out of 748 fulfill the criteria for probable YSOs, leading to a contamination of $\sim 9 \pm 1\%$. Under the assumption of a contamination that is high sample-wide, 790 out of our 8\,781 objects are contaminants.

We also used two representative areas with the size of $15' \times 15'$, located well outside the clouds, i.e. in areas where one would not expect the presence of YSOs, to estimate the contamination of luminous background objects. The total combined number of infrared excess sources in the two areas are 27 objects. Comparing those two areas to the total combined survey area of $5.30$~deg$^2$, the number of overall background contaminants can be estimated as 1\,145 $(\sim 13 \pm 0.4\% $ of the total YSO candidate population). The results of \citet{Oliveira_2009,Oliveira_10} can be interpreted as an extreme case.

The possible contamination of the photometry due to diffuse nebula emission is in most areas rather weak in the NIR and \textit{Spitzer} $4.5\,\mu$m map. As already described in Sect. \ref{subsec:combining_VISTA_SPITZER}, we used the IRAC~2 band for our analysis since it is least affected by PAH emission and, therefore, minimizes the contamination of the photometry of our stars.

The contamination of about 10--15\% used for our further analysis takes the determined contaminations of  $\sim 9 \pm 1\%$ and $\sim 13 \pm$ 0.4$\%$ into
account. Under these conditions, as well as in the extreme case of \citet{Oliveira_2009,Oliveira_10}, the statistical analysis of the spatial distribution of the excess-selected YSOs is still valid since we are not aiming at a reliable identification of individual YSOs.

\section{The spatial distribution of the young stars}
\label{sec:spatial_dist}

\begin{figure}
\resizebox{\hsize}{!}{\includegraphics{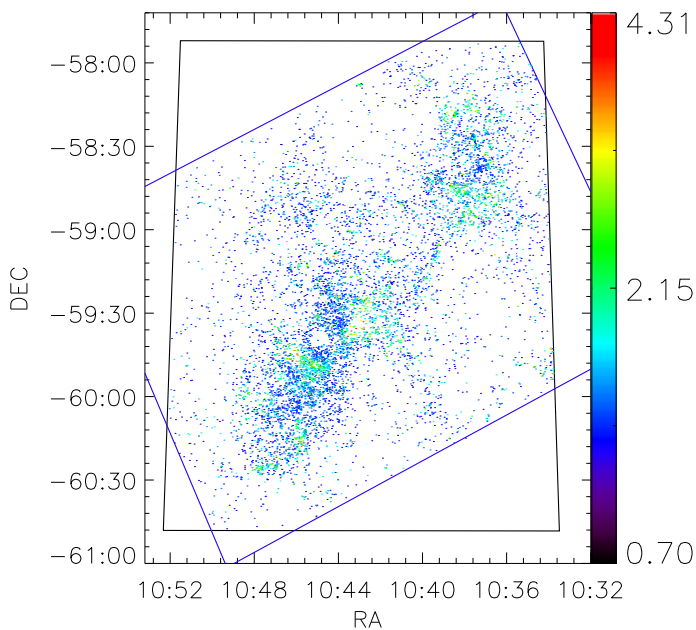}}
\caption{Spatial distribution of the 8\,781 selected infrared excess sources. The color of the individual point shows the $(J-H)$ color of each object. The white hole in the central region with a single spot in the center marks the
region around $\eta$~Car, where the catalogs list no sources due to very strong saturation effects in the \textit{Spitzer} and VISTA images. The black box represents the VCNS area, the blue box the \textit{Spitzer} survey area.}
\label{fig:J-H_K-IRAC2_map_07_45}
\end{figure}

In the map of the spatial distribution of the MIR-excess-selected cYSOs (Fig. \ref{fig:J-H_K-IRAC2_map_07_45}), the surface density of cYSOs drops toward the boundaries of the investigated region. This suggests that the (very conservative) selection criteria we have used actually shows the young stellar population related to the Carina nebula and is no longer dominated by background contamination.

\subsection{Determination of the surface densities}

To investigate the spatial distribution of the MIR-excess-selected cYSOs in the Carina nebula, we performed a nearest-neighbor analysis of the surface density. For each cYSO, we computed the angular distance to the $n^{th}$ nearest neighbor, $r_n$. These values are then used to compute an estimate of the local surface density of cYSOs, the so-called surface density estimator $\mu_n$ \citep{Casertano_Hut_nn} with the formula

\begin{equation}
\label{density}
\mu_n=\frac{n-1}{\pi r_n^2}.
\end{equation}

\noindent We used the value $n = 20$ in our analysis.

\begin{figure*}
\centering
\includegraphics[width=17cm]{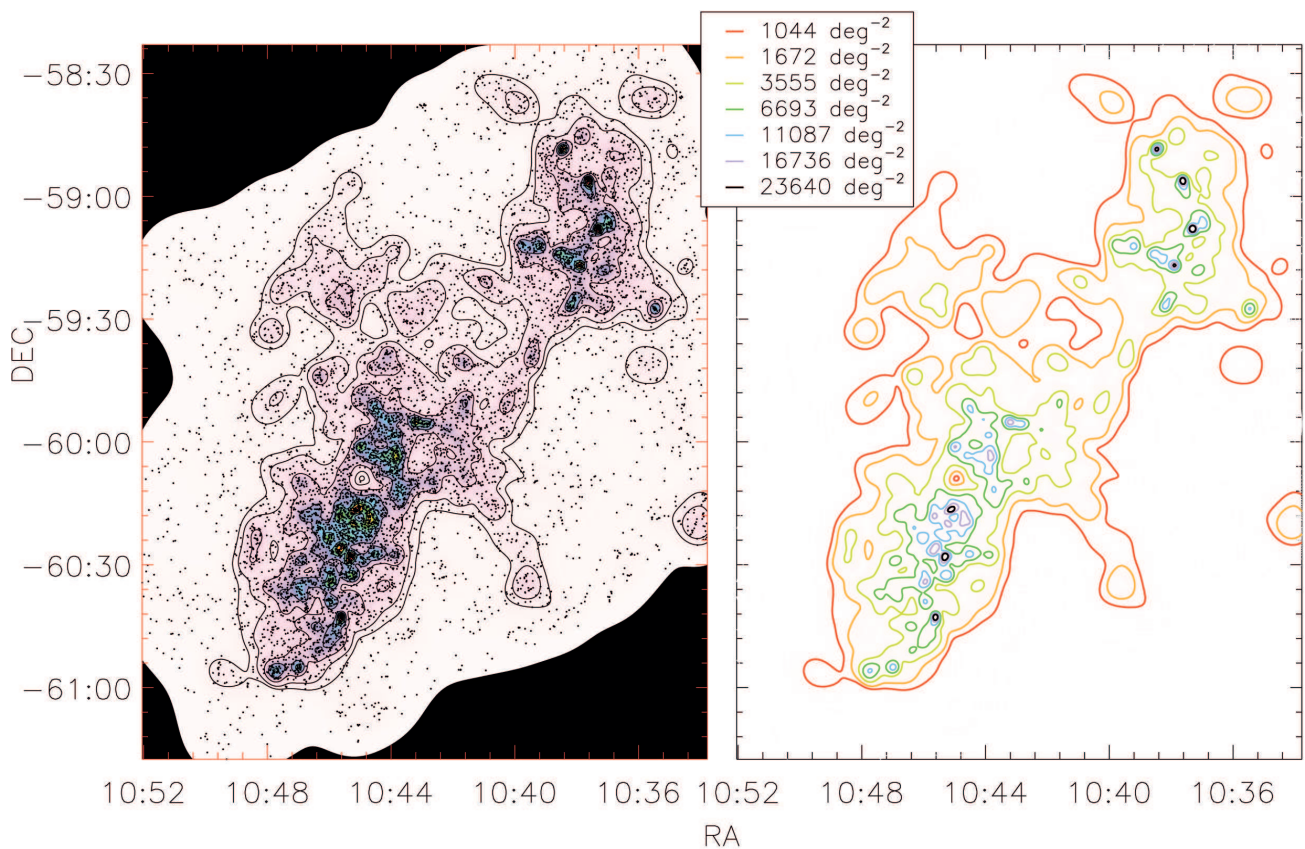}
\caption{\textbf{Left:} Surface density of the infrared-excess-selected cYSOs resulting from the nearest neighbor analysis. The black dots represent the individual 8\,781 cYSOs. The locations of the main clusters like Tr~14, Tr~15, Tr~16, and NGC~3324 are easily visible. The black solid area is not covered by both surveys and therefore the density drops down to zero. The curvy border is a result of the contour smoothing. \textbf{Right:} Contour representation of the surface density of the infrared-excess-selected cYSOs. The contours show surface density values from $\mu_{\text{min}}=1\,044~\text{deg}^{-2}$ (red) to $\mu_{\text{max}}=23\,640~\text{deg}^{-2}$ (black), with the levels $\mu_{k}=(k/6)^2 \cdot (\mu_{\text{max}}-\mu_{\text{min}}) + \mu_{\text{min}}$, with $k=[0,6]$.}
\label{fig:nn_20}
\end{figure*}

To construct a map of surface density values on a regular grid, we used a Delaunay triangulation\footnote{Points in the $\mathbb{R}^2$  get connected to triangles in such a way that no other point is lying within the circles on which the corners of the triangles are located \citep{Klein2005_Delauney}.}. The triangulated grid has $X \times Y = 636 \times 1\,000$ ($\Delta \alpha \times \Delta \delta = 16.4'' \times 16.4''$) points. The resulting map is shown in Fig. \ref{fig:nn_20}. The structure and especially the strong decrease in the surface density of the excess sources, which is already indicated in the infrared excess source position map (Fig. \ref{fig:J-H_K-IRAC2_map_07_45}), becomes more visible. The well known major clusters Tr~14, Tr~15, Tr~16, and NGC~3324 (for comparison see Fig. \ref{fig:survey_area}) are also easily visible as strong peaks in the local surface density.

With a surface density threshold of 1\,044~deg$^{-2}$ (lowest contour of our nearest-neighbor analysis) an area of 2.11~deg$^2$ was determined. Within this area (see Fig. \ref{fig:nn_20}) there are 7\,271 of our infrared excess-selected objects ($\sim83\%$). The area covered simultaneously by the \textit{Spitzer} IRAC and VISTA surveys is $\sim5.30$~deg$^2$.  Around $83\%$ of our infrared excess-selected objects are located in $\sim40\%$ of the fully observed area.

We tried different values of $n$. For $n = 15,$ it is possible to better distinguish the different small-scale structures in the central part of the nebula, but on the other hand, fluctuations in the areas with low surface density were detected as clustering (see Sect. \ref{subsect:ident_cluster}). For $n=25,$ the determined area by the density of the infrared-excess-selected objects would be 1.95~deg$^2$ and only 7.6\% smaller than for $n=20$. This shows that our method is quite robust, considering that we increased the number of counting neighbors by five. Nevertheless, for $n=25$ only very few of the known clusters in the center of the CNC were still distinguishable, and therefore we considered this number as too high for our purpose.

\begin{figure}
\resizebox{\hsize}{!}{\includegraphics{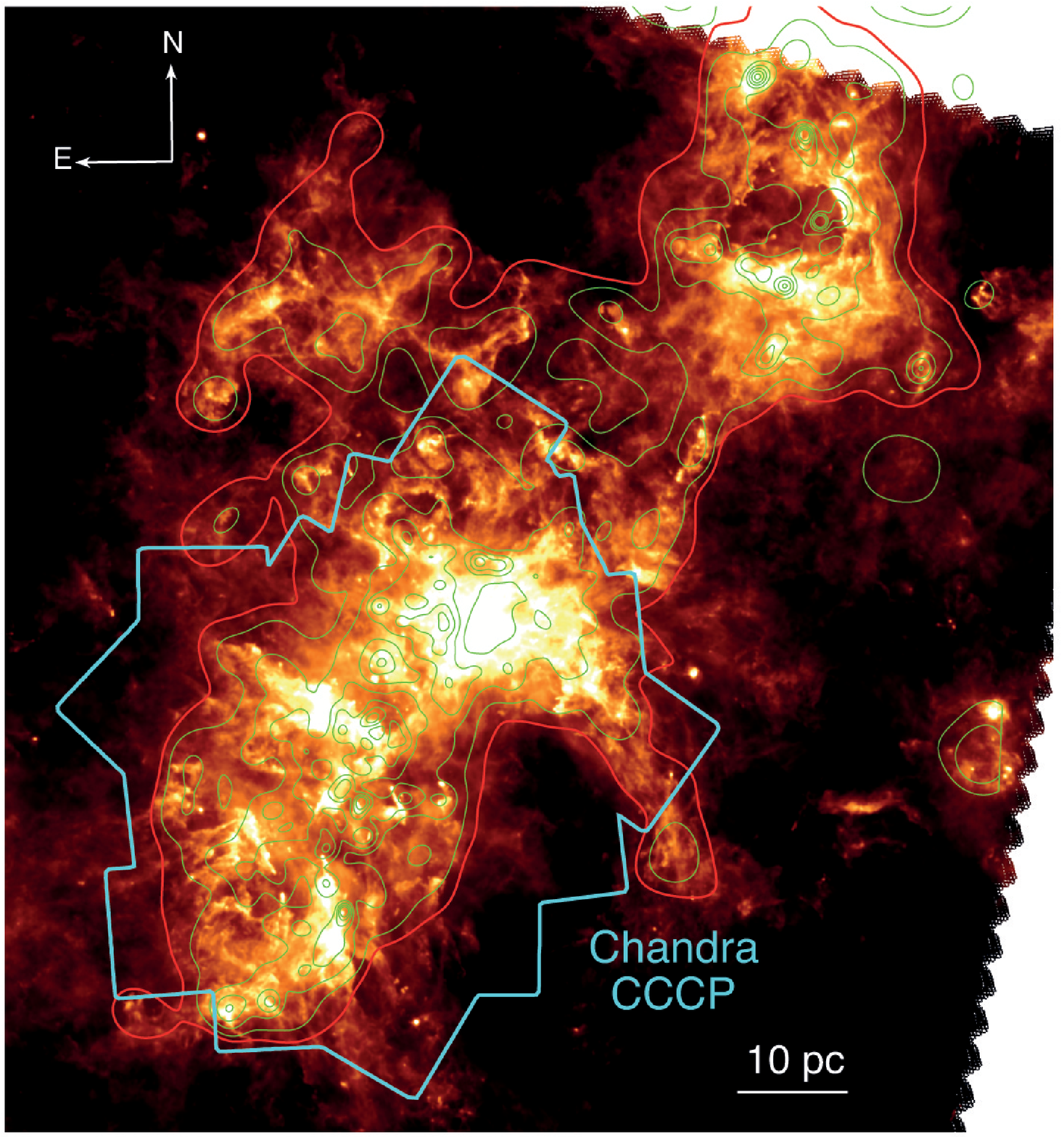}}
\caption{Contours of the surface density distribution of the infrared-excess-selected cYSOs (red) is over-plotted on the \textit{Herschel} $250~\mu$m map. The contour levels are the same as in Fig. \ref{fig:nn_20}. The red contour corresponds to a density of $\mu_{min} = 1 044$~deg$^{-2}$. The cyan polygon marks the CCCP X-ray survey area \citep{Townsley2011CCCP}.}
\label{fig:Carina_boundary}
\end{figure}

Figure \ref{fig:Carina_boundary} shows the surface density of the excess selected cYSOs plotted over the \textit{Herschel} $250~\mu$m map of the clouds \citep{Preibisch2012Herschel1}. The surface density contours for the cYSOs agree very closely with those of the cloud density. This agrees well with the expectation for the spatial distribution of young stars, since one would expect the large majority of the young stars to be located within or very close to the clouds in which they were born. Since the typical values for the velocity dispersion of young stars in star forming regions is $\sim2$~--~3~km/s, and most of the MIR excess-selected cYSOs should have ages of $\lesssim3$~Myr, these stars should not have traveled much more than $\lesssim 3 \times [2$~--~$3] \sim [5$~--~$10]$~pc ($\lesssim 0.25$~deg) away from the birth places. Our confirmation of this expectation suggests that our infrared excess-selected sample of YSO candidates is actually mainly composed of young stars in the Carina nebula, with a low remaining degree of background contamination of about the above-mentioned $\sim 10$--15\%.

The \textit{Herschel} maps (see Fig. \ref{fig:Carina_boundary}) show that the Gum~31 region around the stellar cluster NGC~3324 seems to be connected to the central region of the Carina nebula. The \ion{CO}{} maps of \citet{Yonekura2005CO} revealed similar radial velocities ($V_{LSR}=-24 \ldots -21$~km\,s$^{-1}$) of the clouds surrounding Gum~31 and the central part of Carina ($V_{LSR}=-26 \ldots -18$~km\,s$^{-1}$). A common distance of Gum~31 and the Carina nebula is also suggested by the results of \citet{Walborn_82}, who found that the distance
modulus values of the brightest stars in NGC 3324 are consistent with those of the clusters Tr~16/Col~228 in the central parts of the Carina nebula. All these are clear hints that Gum~31 and the Carina nebula are spatially connected and have the same distance.

\subsection{Identification of young star clusters}
\label{subsect:ident_cluster}
\begin{figure}
\resizebox{\hsize}{!}{\includegraphics{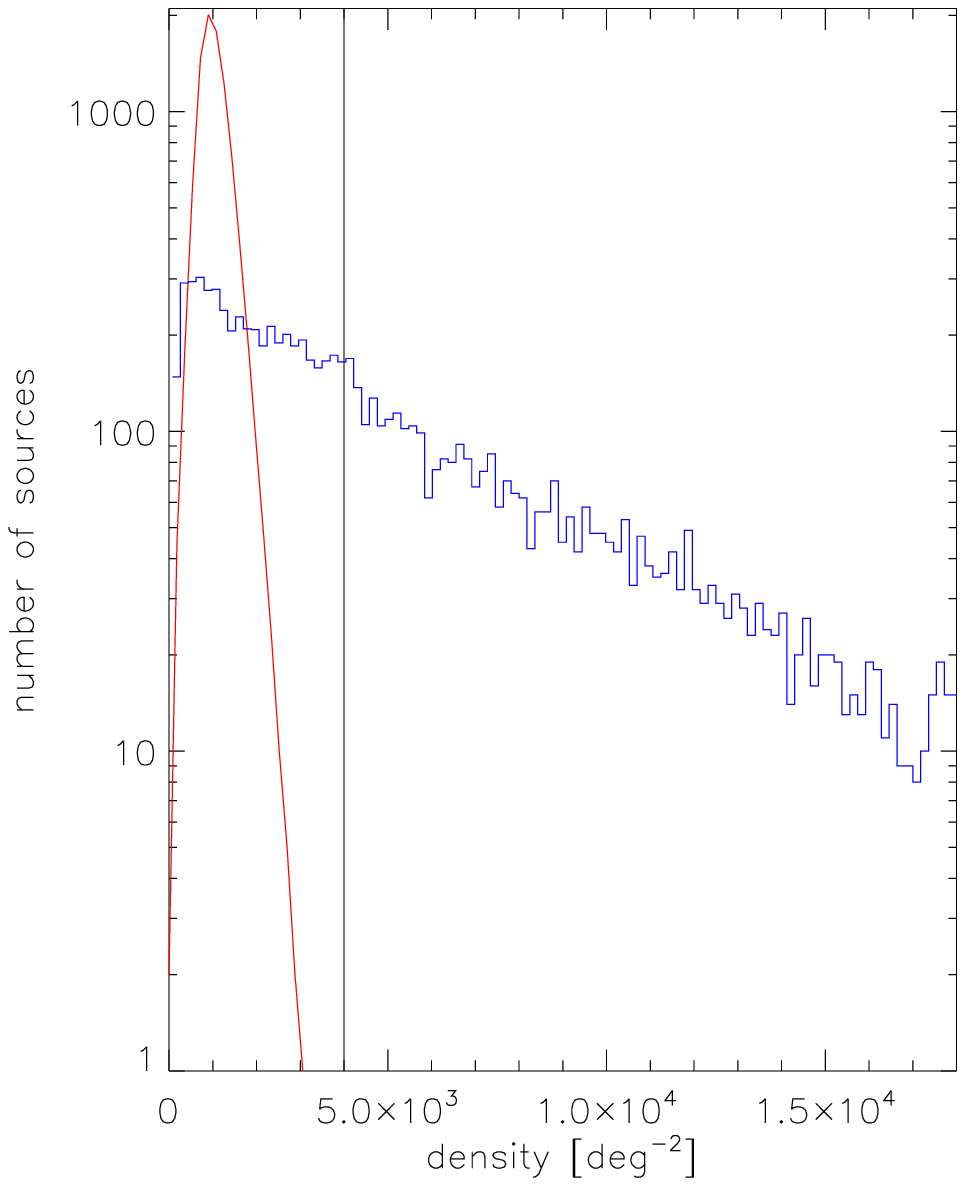}}
\caption{Histogram of the nearest-neighbor analysis surface densities for the infrared-excess-selected sources. The red curve visualizes the surface density of the same number of randomly distributed stars. The vertical line indicates our cut-off limit for clusters at 4\,000~deg$^{-2}$.}
\label{fig:density_histogram}
\end{figure}

The nearest-neighbor analysis can also be used to find new clusters of young stars in the CNC. To determine the density threshold above which we consider a group of cYSOs to be a “cluster”, we first inspected the distribution of surface density estimator values (for n=20) of all cYSOs and compared it to the expected distribution of an equally large number of point sources that are randomly drawn in position from a uniform distribution. In Fig. \ref{fig:density_histogram}, the observed distribution of cYSOs surface densities is compared to the expected density distribution for an equal number of randomly drawn points from a uniform distribution \citep{Chandrasekhar_1943}. The two distributions deviate considerably. The excess of high surface density values for the cYSOs over the expected number of high surface density values is the imprint of the clusters.

\begin{figure*}
\includegraphics[width=17cm]{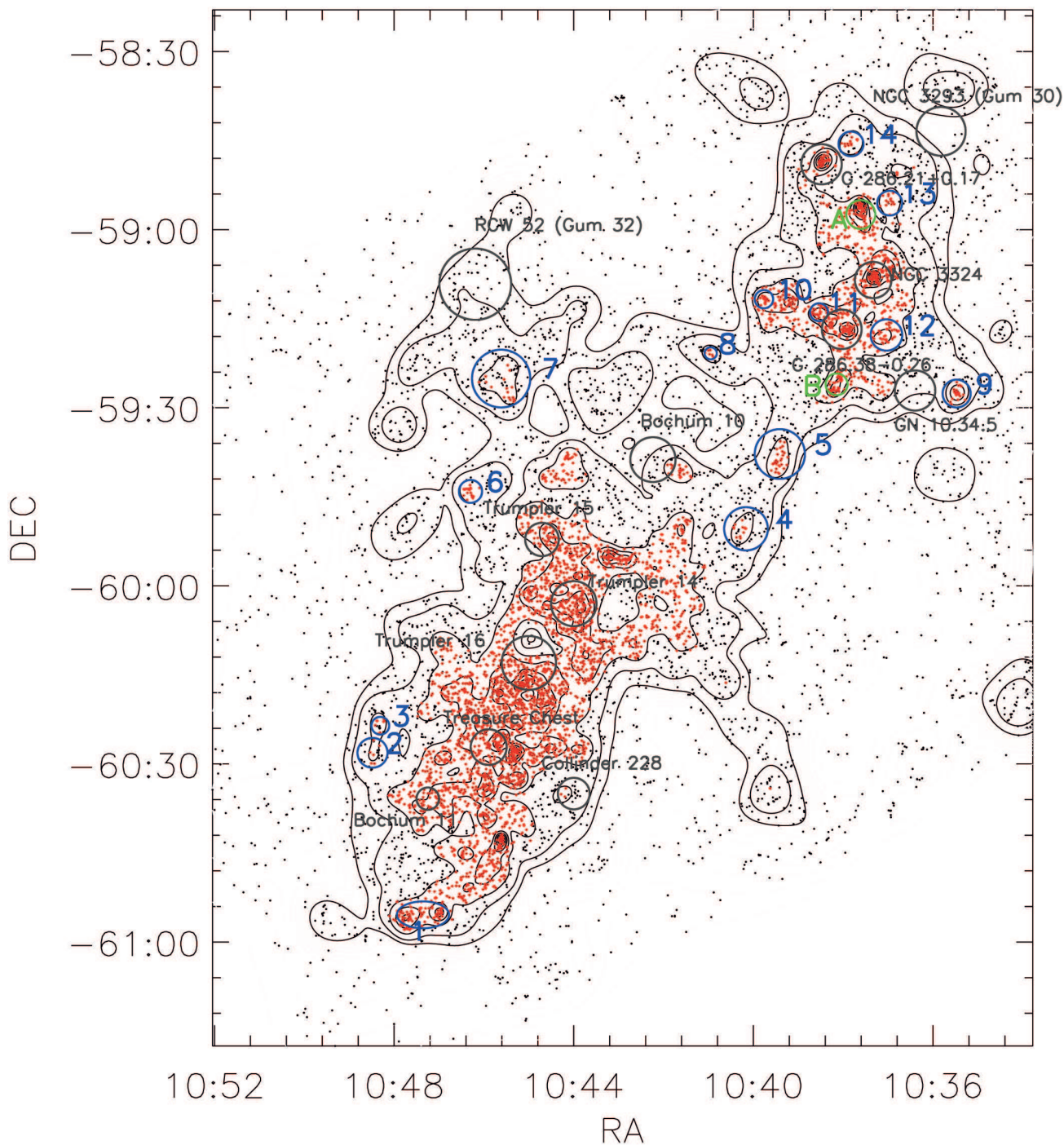}
\caption{Map of all infrared-excess-selected sources. The red dots represent all sources for which the density is higher than the cut-off density (4\,000~deg$^{-2}$). The black circles mark the previously known clusters, while blue circles represent the newly identified clusters. The derived details about these clusters, as well as their coordinates, are listed in Table \ref{tab:new_clusters}.}
\label{fig:new_cluster}
\end{figure*}

To identify new clusters, we chose a conservative surface density threshold of $\mu \ge 4\,000$~deg$^{-2}$ in order to avoid false cluster detection. On physical scales, this corresponds to a cYSOs surface density of 2.48~pc$^{-2}$. This is a strict lower limit to the true surface density of stars in the clusters for two reasons: First, the very strict selection criteria we applied to keep the background contamination as low as possible exclude a substantial fraction of the TSOs with weaker excesses. Second, our infrared-selected sample of cYSOs is known only to be complete down to $0.5~M_\odot$. Therefore, our sample is a rather conservative lower limit to the full number of YSOs in the region. In Fig.~\ref{fig:new_cluster}, all cYSOs with a local surface density above our limit are indicated in red. One can see that the whole central region around the clusters Tr~14, Tr~15, and Tr~16, as well as the region around NGC~3324, have considerably higher densities. Since these central regions have already been well studied so that the existing stellar clusters are already known \citep[see, e.g.,][]{Smith2010South_Pillars}, we restrict our analysis to the more peripheral regions of the complex. Table \ref{tab:new_clusters} lists the resulting 14 identified clusters.

In the following we present an overview of the newly discovered clusters and discuss their relation to the cloud structure of the CNC. We investigated whether the clustered stars follow the cloud structure in the \textit{Herschel} FIR maps \citep{Gaczkowski2012Herschel2} for the wavelengths of $70~\mu$m, $160~\mu$m, $250~\mu$m, $350~\mu$m, and $500~\mu$m. All estimated spatial sizes and the members of the 14 clusters are plotted over the \textit{Herschel} maps in Fig. \ref{fig:cluster_Herschel}. For cluster 14, only data of the $70~\mu$m and $160~\mu$m bands are available.

Cluster 1 lies at the edges of the very dense central region of the Carina nebula. With the detection method and threshold used in this analysis, it is hard to distinguish whether this cluster is one large elongated cluster or if there are actually two clusters. The clusters 4--7, 10, 12, and 14 are clearly located in the dusty clouds visible in the \textit{Herschel} maps. Also the surface density structure of the distribution of the stars is very analogous to that of the cloud structure. In clusters 2, 3, and 8, the stars are located at the outer rims of the clouds. After comparing the VISTA and \textit{Herschel} images, we find that they are most probably still embedded. Comparing the cloud structure in the \textit{Herschel} images, Clusters 2 and 3 could also be connected. Cluster 9 is surrounded by a ring-like cloud structure as seen in the \textit{Herschel} maps. Most of the stars in this cluster are found somewhere inside the ring. The others are located directly on the clouds and were maybe born in that ring-like structure. Clusters 11 and 13 are not associated with clouds.

\begin{table*}
\caption{Identified clusters with their central coordinates, the radius, and a lower limit to the density is given based on the surface density plot of Fig. \ref{fig:nn_20}. Clusters 1 and 12 are already listed in Simbad as SPW2010 and G286.2566--00.3236, respectively. The radii of Cluster 1 give the minor  and the major axes owing to its elliptical shape.}
\label{tab:new_clusters}
\centering
\begin{tabular}{rrrrrr}
\hline
\hline
\multicolumn{1}{c}{name} & \multicolumn{1}{c}{\#} & \multicolumn{1}{c}{RA (J2000)} & \multicolumn{1}{c}{DEC (J2000)} & \multicolumn{1}{c}{radius [arcmin]} & \multicolumn{1}{c}{$\mu_{max}$ $[\text{pc}^{-2}]$}\\
\hline
1& J104721-602524 & 10:47:21.4 & $-$60:25:24  & 2.3/4.4  &      $\ge 4.15$  \\ 
2& J104830-595858  & 10:48:30.8 & $-$59:58:58  & 1.2            &       $\ge 2.21$  \\ 
3& J104818-595333 & 10:48:18.4 & $-$59:53:33  & 2.3             &       $\ge 2.21$  \\ 
4& J104009-591903 & 10:40:09.9 & $-$59:19:03  & 2.5     &       $\ge 2.21$  \\ 
5& J103924-590721 & 10:39:24.7 & $-$59:07:21  & 4.2     &       $\ge 2.21$  \\ 
6& J104617-591405 & 10:46:17.8 & $-$59:14:05  & 1.9     &       $\ge 2.21$  \\ 
7& J104537-585505 & 10:45:37.0 & $-$58:55:05  & 4.9     &       $\ge 2.21$  \\ 
8& J104057-585050 & 10:40:57.1 & $-$58:50:50  & 1.0             &       $\ge 2.21$  \\ 
9& J103528-585736 & 10:35:28.4 & $-$58:57:36  & 2.4     &       $\ge 4.15$  \\ 
10& J103945-584143 & 10:39:45.9 & $-$58:41:43  & 1.5            &       $\ge 4.15$  \\ 
11& J103832-584355 & 10:38:32.2 & $-$58:43:55  & 1.5            &       $\ge 6.88$  \\ 
12& J103702-584745 & 10:37:02.2 & $-$58:47:45  & 2.7    &       $\ge 4.15$  \\ 
13& J103657-582531 & 10:36:57.8 & $-$58:25:31  & 2.1    &       $\ge 2.21$  \\ 
14& J103749-581533 & 10:37:49.5 & $-$58:15:33   & 2.1   &       $\ge 2.21$  \\
A & J103736-582727 & 10:37:36.5 & $-$58:27:27  & 2.4    &       $\ge 14.67$ \\ 
B & J103808-585603 & 10:38:08.7 & $-$58:56:03  & 1.8    &       $\ge 6.88$  \\  
\hline 
\end{tabular}
\end{table*}

\subsection{Portion of cYSOs in clusters}

The surface density of the cYSOs allows us to estimate how many of these objects are located in clusters and what percentage of them constitutes a more distributed population. Using the surface density threshold of 4\,000~deg$^{-2}$ defined above, we find that 4\,137 of the 8\,781 cYSOs, i.e. 47\%, are located in clusters. This suggests that roughly half of the cYSOs constitute a non-clustered, dispersed population.

This result agrees with the results on the spatial distribution of the X-ray selected YSOs in the CCCP \citep{Feigelson2011Carina_Xray,Preibisch2011bNIR_XRAY}, which established that about half of the young stars are members of clusters or stellar groups, while the other half constitutes the “widely-distributed population”, which is spread out through the entire CCCP area. Our findings extend this result from the 1.4~square-degree CCCP field to the full area of the CNC.

This highly populous distributed population of YSOs supports the classification of the Carina Nebula Complex as a stellar association. The prominent stellar clusters Tr~14, Tr~15, and Tr~16 in the central region and NGC~3324 actually contain just about (or less than) half of the total young stellar population. This spatial configuration is in good agreement with theoretical models of the stochastic dynamical evolution of stellar complexes with an extended history of star formation \citep[see, e.g.:][]{Parker_2014}.

\section{An estimate of the full young stellar population in the CNC}
\label{sec:YSO}

Based on the size of the X-ray selected YSO sample, \citet{Feigelson2011Carina_Xray} estimated that the total stellar population (i.e., down to stellar masses of $\sim0.1~M_\odot$) in the 1.4~square-degree CCCP field should be $\sim104\,000$. We can use this to obtain an estimate for the total number of YSOs in the full area of the CNC.

Of our 8\,781 infrared-excess-selected cYSOs, 4\,618 are located in the CCCP field. The ratio of this number to the expected total number of YSOs shows that our cYSOs sample represents about 5\% of the full young stellar population in this area. If we assume that the ratio between MIR excess-selected and X-ray-selected young stars is about the same for the areas outside the CCCP field, we can estimate that the total extrapolated number of young stars in the entire CNC, including the region NGC~3324, is thus 164\,000.

\section{Summary and conclusions}
\label{sec:summary}

The combination of the NIR data ($J$,$ H$, and $K_s$-band)  of the VCNS with the \textit{Spitzer} MIR data ($3.6~\mu$m, $4.5~\mu$m, $5.8~\mu$m, and $8.0~\mu$m) \citep{Ohlendorf2013GUM31} revealed new insight into the spatial distribution and the clustering properties of the young stellar populations in the CNC. The area covered simultaneously by both surveys is $\sim5.30$~deg$^2$. The extension to the MIR helped toward a better selection of infrared-excess-selected objects and a reduction of the background contamination.

In total, 8\,781 infrared excess sources were selected based on the following criteria: $J-H > 0.70$~mag and $K_s-[4.5] > 0.49$~mag, a location below the reddening band with a slope of 2.27, and a S/N>10. This is a very conservative (and thus necessarily incomplete), but therefore fairly clean sample of cYSOs, with a low level $(\sim10$--15\%) of remaining background contamination. This sample allowed us, for the first time, to determine the boundaries of the spatial distribution of young stars associated with the Carina nebula. With a nearest-neighbor analysis, we were able to determine the boundaries of the young stellar population in the CNC, where the spatial density of infrared-excess-selected sources drops to a limiting lower density of 1\,044~deg$^{-2}$ (0.65~pc$^{-2}$).

Within this area of 2.11~deg$^2$ (3\,400~pc$^2$) (see Fig. \ref{fig:nn_20}), there are 7\,271 of our infrared-excess-selected objects . Therefore, $\sim83\%$ of all infrared-excess-selected YSO candidates are located in $\sim40\%$ of the full observed area. Of these 7\,271 objects, 4\,137 are in clusters. This suggests that roughly half of the cYSOs constitute a non-clustered, dispersed population.
Most of these newly identified clusters appear to be well connected to the dusty clouds visible in the \textit{Herschel} FIR maps \citep{Preibisch2012Herschel1}.

Based on an estimate of the total young stellar population in the area of the CCCP X-ray survey \citep{Feigelson2011Carina_Xray}, we can estimate a total number of $\sim164\,000$ YSOs for the full area of the Carina nebula complex that includes the region NGC~3324.

\begin{acknowledgements}
This work is based in part on observations made with the \textit{Spitzer} Space Telescope, which is operated by the Jet Propulsion Laboratory, California Institute of Technology under a contract with NASA. This work was supported by funding from Deutsche Forschungsgemeinschaft under DFG project number PR 569/9-1. Additional support came from funds from the Munich Cluster of Excellence “Origin and Structure of the Universe”.
\end{acknowledgements}

\bibliographystyle{aa}
\bibliography{bibliography}

\appendix
\section{FIR images of the identified clusters}

\begin{figure*}
\centering
\begin{minipage}{1.0\textwidth}
\includegraphics[width=17cm]{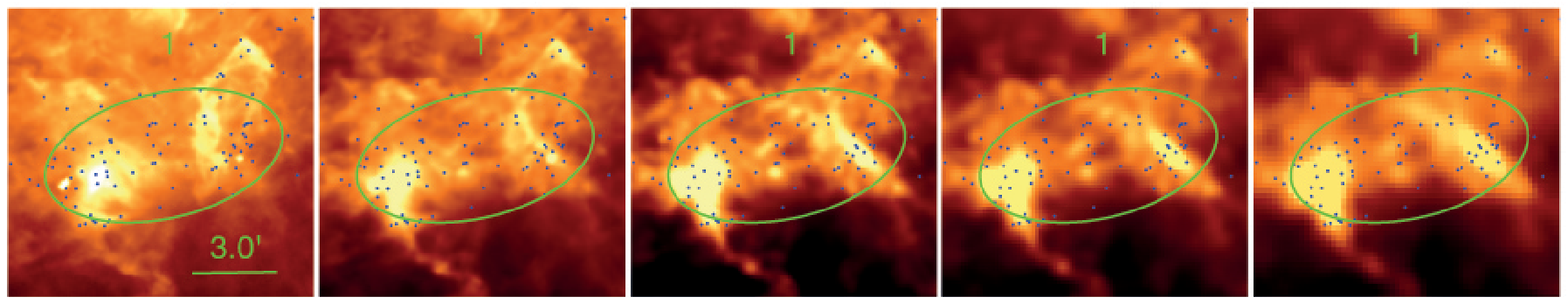}
\end{minipage}

\begin{minipage}{1.0\textwidth}
\includegraphics[width=17cm]{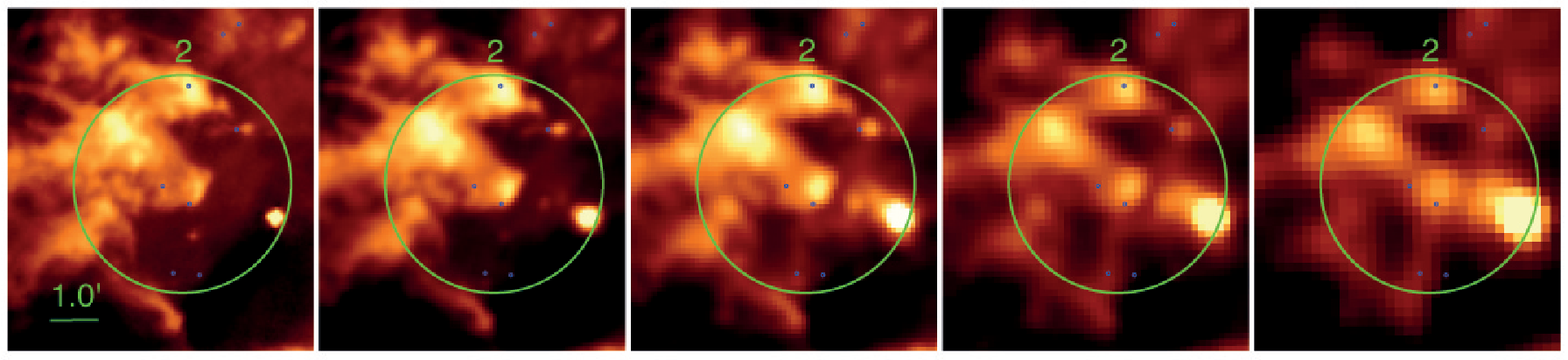}
\end{minipage}

\begin{minipage}{1.0\textwidth}
\includegraphics[width=17cm]{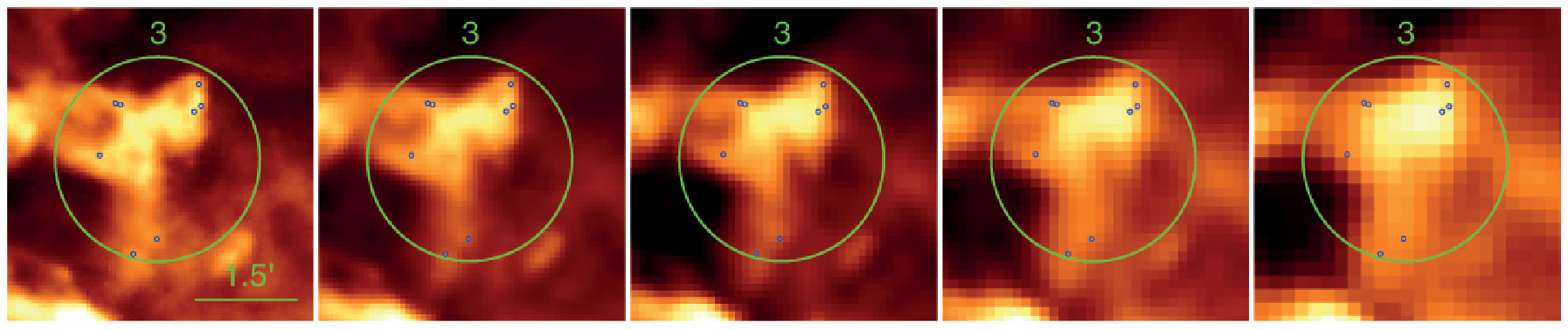}
\end{minipage}

\begin{minipage}{1.0\textwidth}
\includegraphics[width=17cm]{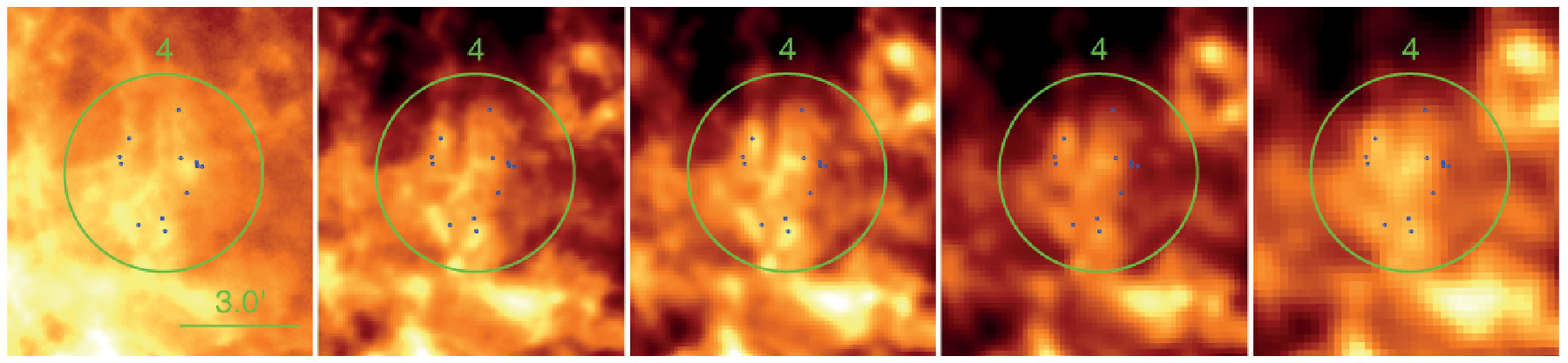}
\end{minipage}

\begin{minipage}{1.0\textwidth}
\includegraphics[width=17cm]{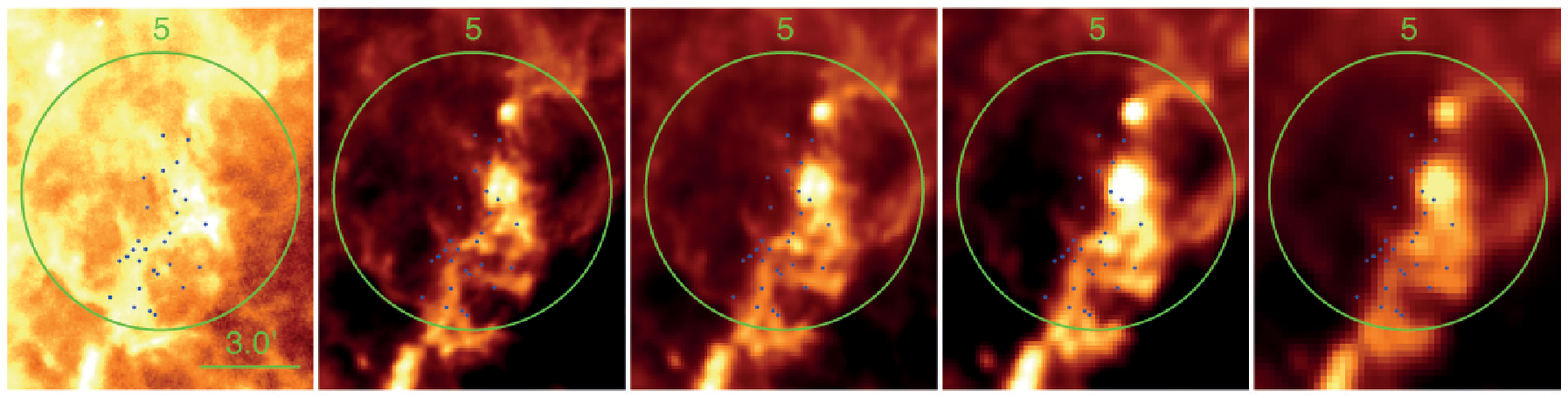}
\end{minipage}

\begin{minipage}{1.0\textwidth}
\includegraphics[width=17cm]{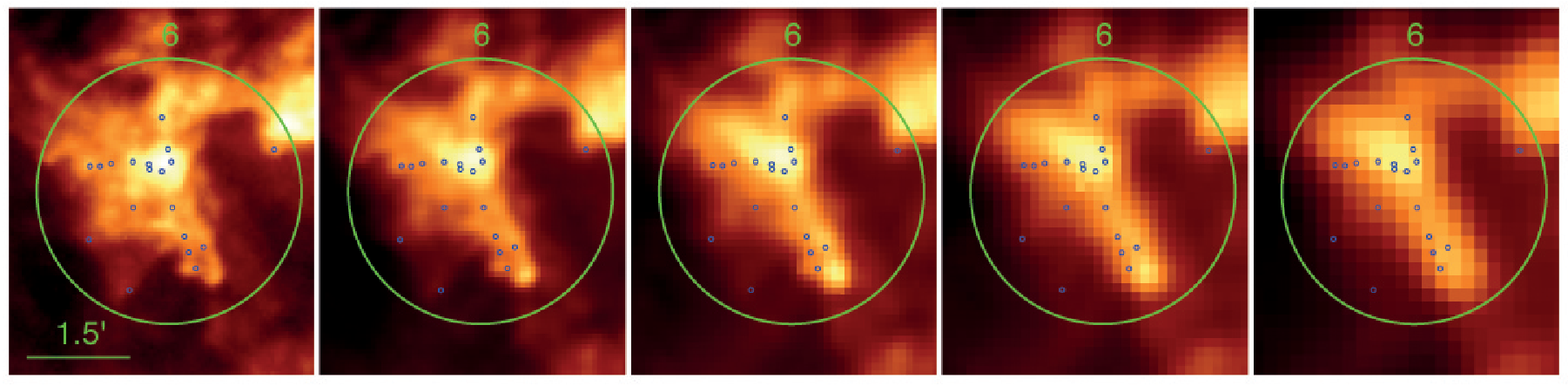}
\end{minipage}
\caption{\textit{Herschel} FIR maps of the regions around the identified clusters of infrared excess source. From left to right: $70~\mu$m, $160~\mu$m, $250~\mu$m, $350~\mu$m, and $500~\mu$m. For cluster 14 only the $70~\mu$m and $160~\mu$m bands are available.}
\label{fig:cluster_Herschel}
\end{figure*}

\setcounter{figure}{9}

\begin{figure*}
\centering
\begin{minipage}{1.0\textwidth}
\includegraphics[width=17cm]{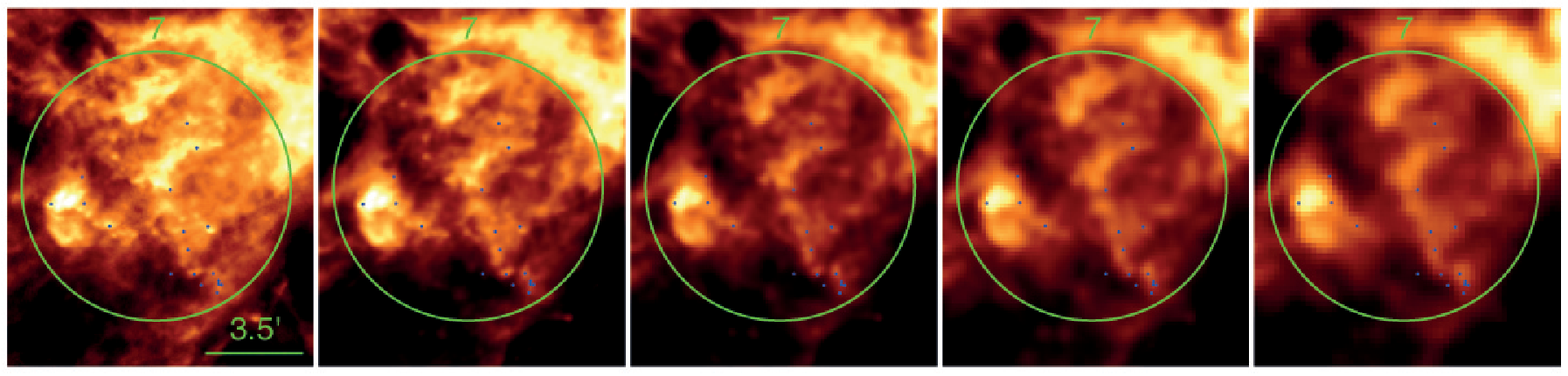}
\end{minipage}

\begin{minipage}{1.0\textwidth}
\includegraphics[width=17cm]{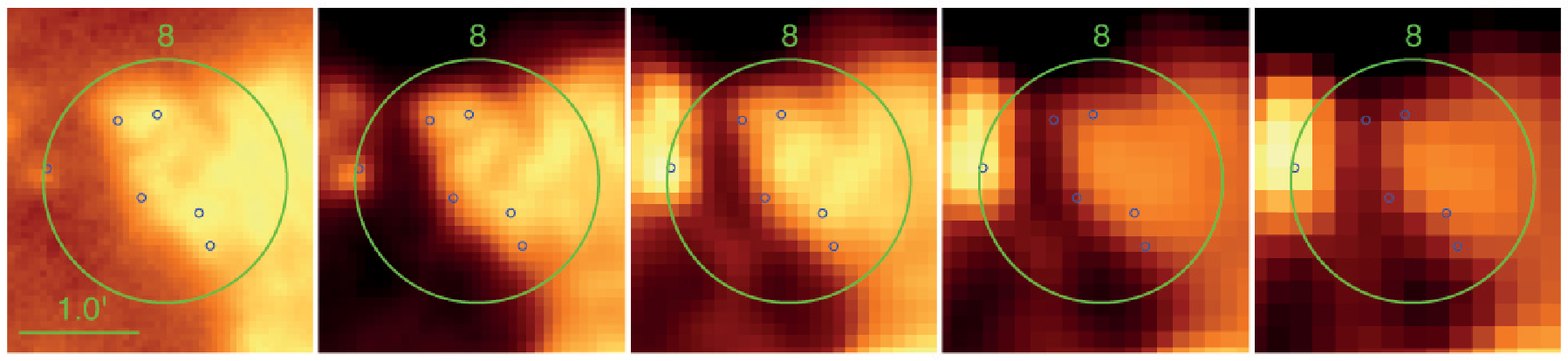}
\end{minipage}

\begin{minipage}{1.0\textwidth}
\includegraphics[width=17cm]{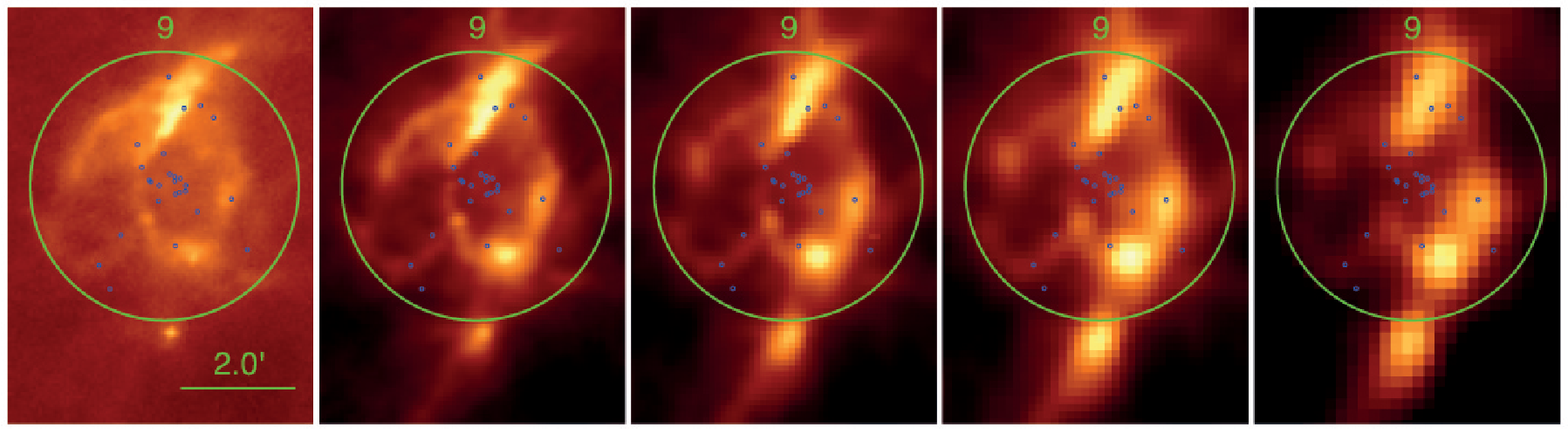}
\end{minipage}

\begin{minipage}{1.0\textwidth}
\includegraphics[width=17cm]{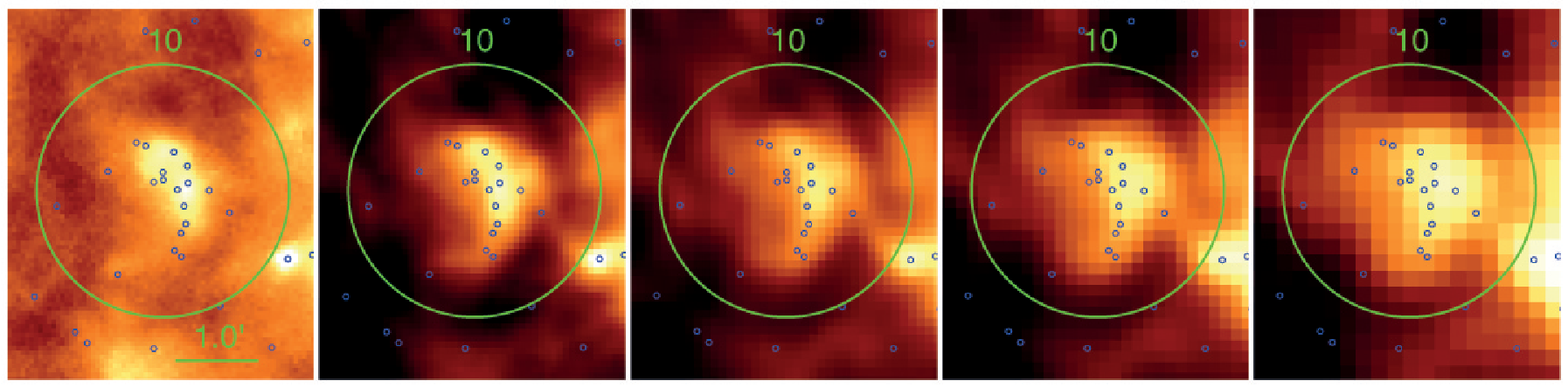}
\end{minipage}

\begin{minipage}{1.0\textwidth}
\includegraphics[width=17cm]{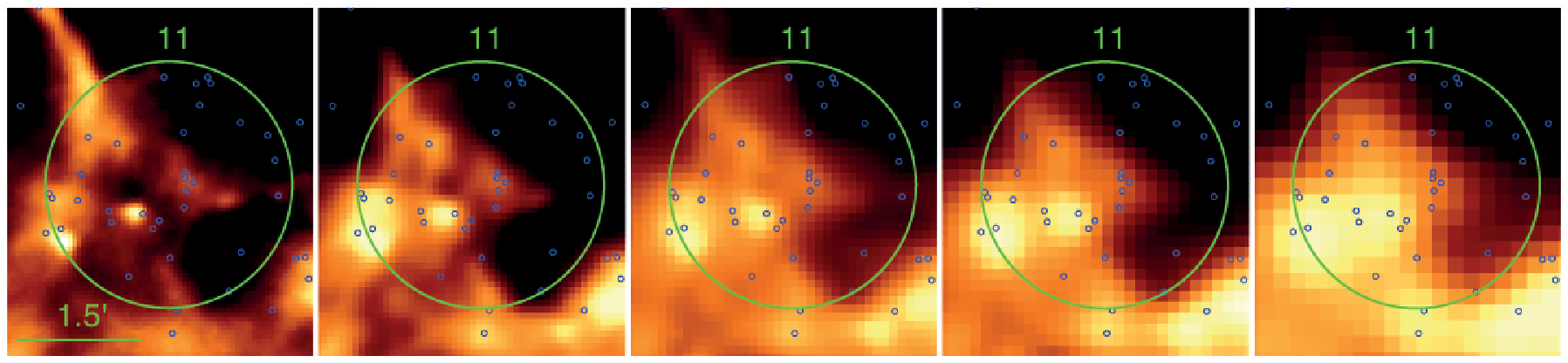}
\end{minipage}

\caption{continued...}
\end{figure*}

\newpage

\setcounter{figure}{9}

\begin{figure*}
\centering
\begin{minipage}{1.0\textwidth}
\includegraphics[width=17cm]{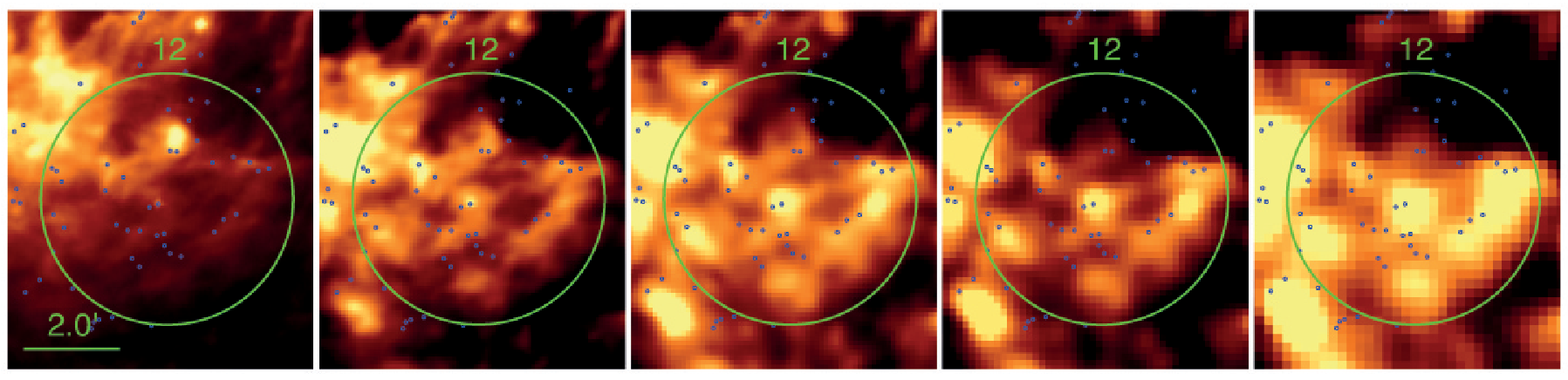}
\end{minipage}

\begin{minipage}{1.0\textwidth}
\includegraphics[width=17cm]{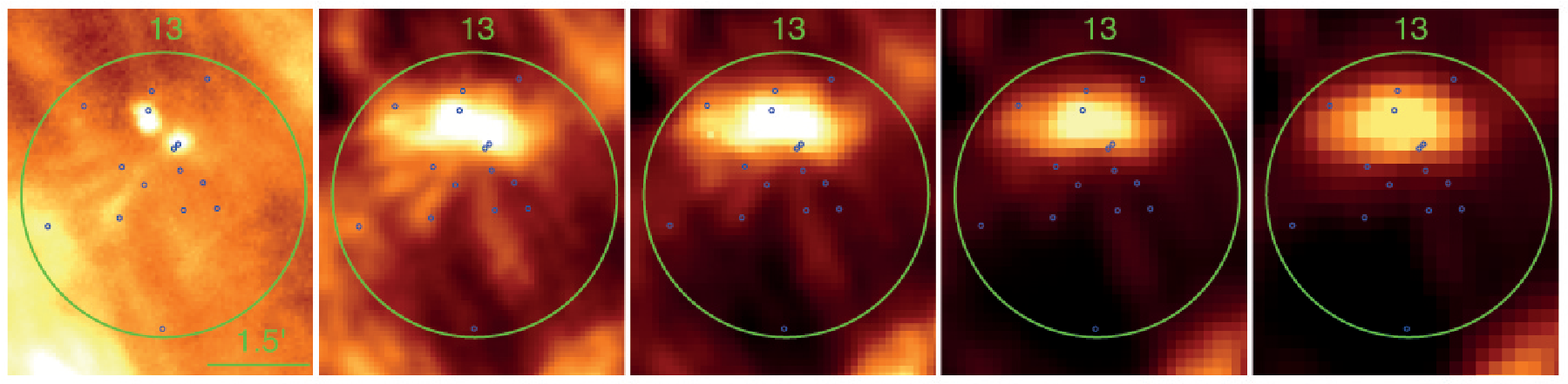}
\end{minipage}

\begin{minipage}{1.0\textwidth}
\includegraphics[width=17cm]{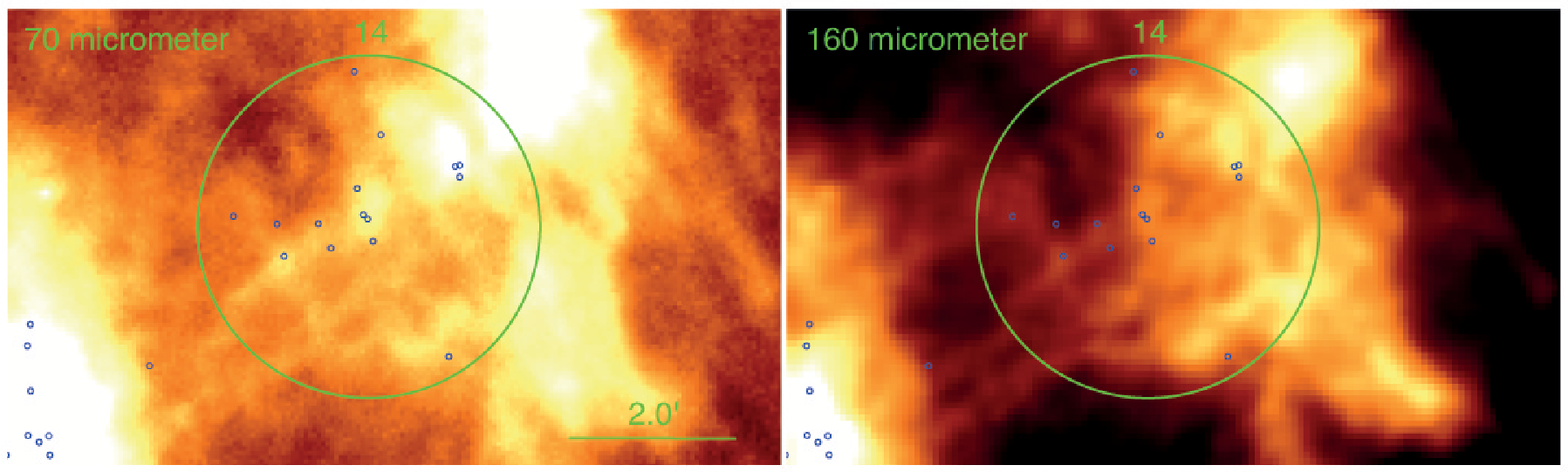}
\end{minipage}

\caption{continued...}
\end{figure*}

\section{The catalog of infrared excess objects in the VCNS survey area}

\begin{sidewaystable*}
\caption{Combined VCNS \textit{Spitzer} photometric infrared excess source catalog}
\label{tab:IR_exces_cat}       % Give a unique label
\begin{tabular}{crrrrrrrrrrrrrr}
VCNS -- J & $J$ & $H$  & $K_s$ & [3.6] & [4.5] & [5.8] & [8.0] & S/N & S/N & S/N  & S/N & S/N & S/N & S/N \\
$[$\,J2000\,$]$ & [mag] & [mag] & [mag] & [mag] & [mag] & [mag] & [mag] & ($F_J$) & ($F_H$) & ($F_{Ks}$) & ($F_{[3.6]}$) & ($F_{[4.5]}$) & ($F_{[5.8]}$) & ($F_{[8.0]}$)  \\
(1)& (2) & (3) & (4) & (5) & (6) & (7) & (8) & (9) & (10) & (11)& (12) & (13) & (14) & (15) \\
\hline
103346.62$-$595137.5 & 17.28 & 15.48 & 15.02 & -9.99 & 13.68 &  -9.99 & -9.99 & 56.4  & 104.2 & 79.7 & -9.9 & 19.8 & -9.9 & -9.9 \\
103347.48$-$595319.2 & 15.83 & 14.18 & 13.09 & 11.37 & 10.58 &    9.64 &   8.61 & 145.7&  222.1 & 241.3 &     278.1 &     341.1 &     143.5 &     157.4 \\
103347.96$-$594953.7 & 18.10 & 16.39 & 15.27 & 13.48 & 12.63 &  -9.99 & 11.22 &  29.9 &  56.0 &  67.9 &      40.9 &      49.2 &      -9.9 &      18.5 \\
103349.07$-$594349.4 & 18.21 & 16.02 & 14.85 & 13.58 & 12.91 &  -9.99 & -9.99 &  28.1 &  72.5 &  89.3 &      41.4 &      40.2 &      -9.9 &      -9.9 \\
103349.36$-$594345.4 & 18.19 & 15.95 & 14.53 & 12.49 & 11.69 &  11.53 & -9.99 &  28.6 &  76.1 & 109.9 &     130.2 &     142.0 &      28.0 &      -9.9 \\
103349.77$-$594242.2 & 15.13 & 14.09 & 13.54 & 12.92 & 12.67 &  11.88 & -9.99 & 212.0 & 233.5 & 194.6 &      69.1 &      54.3 &      14.9 &      -9.9 \\
103349.94$-$595347.3 & 16.29 & 15.07 & 14.34 & 13.55 & 13.29 &  -9.99 & -9.99 & 108.4 & 134.9 & 119.2 &      37.6 &      26.4 &      -9.9 &      -9.9 \\
103350.28$-$595147.2 & 18.84 & 16.87 & 15.83 & 14.11 & 13.30 &  12.42 & -9.99 &  16.3 &  39.0 &  45.1 &      21.4 &      26.4 &      10.1 &      -9.9 \\
    ...             &  ...  &  ... &  ... & ... & ... & ... & ... & ... & ... & ... & ... & ... & ... & ... \\
105202.57$-$600421.5 & 16.54 & 15.36 & 14.82 & 14.26 & 14.12 & -9.99 & -9.99 & 85.0 & 111.7 &  91.9 &  20.3 &  12.9 & -9.9 &  -9.9\\

\hline \\
\multicolumn{15}{l}{\textbf{Note:} Table \ref{tab:IR_exces_cat} is only available in electronic form at the CDS via anonymous ftp to cdsarc.u-strasbg.fr (130.79.128.5) }\\
\multicolumn{15}{l}{or via http://cdsweb.u-strasbg.fr/cgi-bin/qcat?J/A+A/.}\\
\multicolumn{15}{l}{A portion is shown here for guidance regarding its form and content.}\\
\multicolumn{15}{l}{The meaning of the columns is as follows:}\\
\multicolumn{15}{l}{Col.\ (1): Source name, based on the J2000 celestial coordinates.}\\
\multicolumn{15}{l}{Cols.\ (2) to (4): Near-infrared VCNS photometry in the $J$-, $H$-, and $K_s$-bands.}\\
\multicolumn{15}{l}{Cols.\ (5) to (8): Mid-infrared \textit{Spitzer} photometry in the [3.6], [4.5], [5.8], and [8.0] IRAC bands.}\\
\multicolumn{15}{l}{\hspace{2cm}  Objects without photometry in a band are marked by the number $-9.99$.}\\
\multicolumn{15}{l}{Cols.\ (9) to (15): Signal-to-noise ratio of the individual photometric flux measurements in Cols.~(2) to (8).}\\
\end{tabular}
\end{sidewaystable*}

\end{document}